\documentclass[pre,twocolumn,showpacs,superscriptaddress] {revtex4}
\usepackage{graphics}
\newcommand{\cu}
{\affiliation{Department of Physics, University of Calcutta,
92 Acharya Prafulla Chandra Road, Kolkata 700009, India}}

\begin{document}
  
 \title
{Noise driven dynamic phase transition in a one-dimensional Ising-like model}

\author	{  Parongama Sen }
\cu
\begin{abstract}

The dynamical evolution of a recently  introduced one dimensional model in \cite{biswas-sen} (henceforth referred to as model I), 
has been made stochastic by introducing a parameter 
$\beta$ such that $\beta =0$ corresponds to the  Ising model
and $\beta \to \infty$ to the original model I. The equilibrium
behaviour for any value of $\beta $ is identical: a homogeneous state.  
We argue, from the behaviour of the dynamical exponent $z$, 
 that for any $\beta \neq 0$, the system belongs to the 
dynamical class of model I indicating a dynamic phase transition at 
$\beta = 0$. On the other hand, the persistence 
probabilities in a system of $L$ spins saturate at a value $P_{sat}(\beta, L)  = (\beta/L)^{\alpha}f(\beta)$,
where $\alpha$ remains constant  for all $\beta \neq 0$ supporting the existence
of the dynamic phase transition at $\beta =0$. The scaling function $f(\beta)$  shows a crossover behaviour 
with $f(\beta) = \rm{constant} $ for $\beta <<1$ and $f(\beta) \propto \beta^{-\alpha}$ for $\beta >>1$.

\vskip 0.5cm

\end{abstract}

\maketitle

The effect of noise on  equilibrium behavior is well known, e.g., there are order-disorder phase transitions induced by 
thermal noise observed in many systems.
Noise induced phase transitions may occur in dynamical systems as well when the 
noise can drive the system  from one dynamical class to another. These 
dynamical classes are often characterised by different dynamical exponents.
In this paper, we study  a case of  such a dynamical phase transition in a very simple Ising like spin system.

A dynamical model of Ising spins  has been recently proposed  
in \cite{biswas-sen} (which we refer to as model I henceforth) 
where the state of the spins may change in  two situations:
first when its two
neighbouring domains have opposite polarity,  and in this case
the  spin  orients itself along the 
spins of the neighbouring domain with the  larger size.
This case may arise only when the spin  is at the boundary of the two
domains.
A spin is also flipped when it is sandwiched between two domains of spins
with same sign.
Except for the rare event when  the two neighbouring domains of opposite spins 
are of the same size, the dynamics in the above model is 
deterministic. 
This dynamics leads to a  homogeneous state of either all spin
up or all spin down. Such evolution to absorbing homogeneous 
states are known to occur in systems belonging to
directed percolation (DP) processes, zero temperature Ising model, voter model etc \cite{absorb,vote}.

Model I 
was introduced in the context 
of a social system where the binary 
opinions of individuals are 
 represented by    up and down spin states.
In opinion dynamics models, such representation of opinions by Ising or Potts
spins is quite common \cite{opinion1}. The key feature is the interaction 
of the 
individuals which may lead to phase transitions between a homogeneous state to a heterogeneous state in many cases \cite{opinion2}. 

Model I showed the existence of 
novel  dynamical  behaviour  in a coersening process when compared to the
dynamical behaviour of DP processes, voter model, Ising models etc \cite{hinrich2,stauffer2,sanchez,shukla,derrida}. 
In this work, we have introduced stochasticity in the dynamics of Model I
to see how it affects the coarsening process.

Let $d_{up}$ and $d_{down}$ be the sizes of the two neighbouring
domains of type up and down of a spin at the domain boundary (excluding itself). In model I, probability $P(up)$ that the said spin is up  
is  1 if $d_{up} > d_{down}$, 0.5 if  $d_{up} = d_{down}$ and zero otherwise.
In the simplest possible way to introduce stochasticity, one may 
take the  probability of a boundary spin to be up as  
$P(up) = d_{up}/ (d_{up}+ d_{down})$.
However, there is no parameter
controlling the stochasticity here and moreover, we find that the 
results are identical to the original model I. 

In order to introduce a noise like parameter which can be tuned, we next 
propose that the probability that a spin at the domain boundary is up is 
given by
\begin{equation}
P(up) \propto e^{\beta(d_{up}- d_{down})},
\end{equation}
and it is down with probability 
\begin{equation}
P(down) \propto e^{\beta(d_{down}- d_{up})}.
\end{equation}
The normalised probabilities are  therefore 
$P(up) =  \exp{\beta\Delta}/(\exp(\beta\Delta) + \exp(-\beta\Delta))$ and  
 $P(down)=1-P(up)$,
where $\Delta = (d_{up}- d_{down}) $. 

Obviously, $\beta \to \infty$ cooresponds to  model I while 
letting $\beta =0$ we have  equal probabilities of the up and down states,  making it equivalent to the zero temperature dynamics of the 
nearest neighbour Ising model. Since the equilibrium states for the extreme values $\beta \to \infty$ and 
$\beta =0$  are homogeneous  
 (all up or all down states), it is expected that for all values of $\beta$ they will be remain so  as is indeed the case. 

As far as dynamics is concerned, we investigate 
primarily the 
time dependent behaviour of the order parameter  and
the persistence probability. In the one dimensional chain of length $L$, the order parameter is the conventional
magnetisation given by $M = \frac{|L_{up} - L_{down}|}{L}$
where $L_{up}~~ (L_{down})$ is the number of up (down) spins in the system
and $L = L_{up}+ L_{down}$, the total number of spins.
The average fraction of domain walls  $D_w$, which is the average number of domain walls divided by $L$ is also studied. $D_w$ is identical to the 
inverse of average
domain size. The dynamical evolution of the order parameter and fractaion of domain walls 
is  expected to be governed by the dynamical exponent $z$; $M \propto t^{-1/(2z)}$ and 
$D_w \simeq t^{-z}$ \cite{bray}. 

The persistence probability
of a spin is the probability that it remains in its original state 
upto time $t$ \cite{derrida}. It has been shown to have a power law decay in many systems with 
an associated exponent $\theta$.    
To obtain both the exponents $\theta $ and $z$ in finite systems of dimension $L$ from the 
persistence probability, the following scaling form is often used \cite{pray}
\begin{equation}
P(t,L) = t^{-\theta}f(L/t^{1/z}).
\end{equation}
Another exponent, $\alpha= \theta z$, is associated with the 
saturation value of the persistence probability at $t\to \infty$ when 
$P_{sat}(L) = P(t \to \infty, L) \propto L^{-\alpha}$ \cite{pray}.

In model I, it was numerically obtained that $\theta \simeq	 0.235$ and   $z \simeq 1.0$ giving  
$\alpha \simeq 0.235$,
while in the one dimensional Ising model $\theta = 0.375$ and $z=2.0$ (exact results) giving $\alpha =0.75$.
It is  clearly indicated that  model I and the Ising model belong to two different dynamical classes. 
By introducing the parameter $\beta$ one  can therefore expect a transition from the Ising to the 
model I dynamical behaviour at some specific value of $\beta$. 

With respect to model I, $\beta=0$ is the maximum noise and its inverse may be thought of an effective temperature. On the other hand, from the Ising model viewpoint, 
$\beta$ plays the role of noise.  However it is not equivalent to thermal fluctuations which can affect the state of any spin. 
With $\beta$, flipping of spins  can still occur at the domain boundaries only. Hence, even with this noise, the equilibrium behaviour 
is not disturbed for any value of $\beta$ (even for $\beta \to \infty$ which corresponds to model I) while in contrast, any non-zero temperature can destroy the order of a one dimensional Ising model. 

It is useful to show the snapshots of the evolution of the system over
time for different $\beta$ (Fig.1): to be noted is the fact that 
for any non-zero $\beta$, the system equilibriates very fast compared to the Ising limit $\beta =0$. 

\begin{figure}
\rotatebox{270}{\resizebox*{4.5cm}{!}{\includegraphics{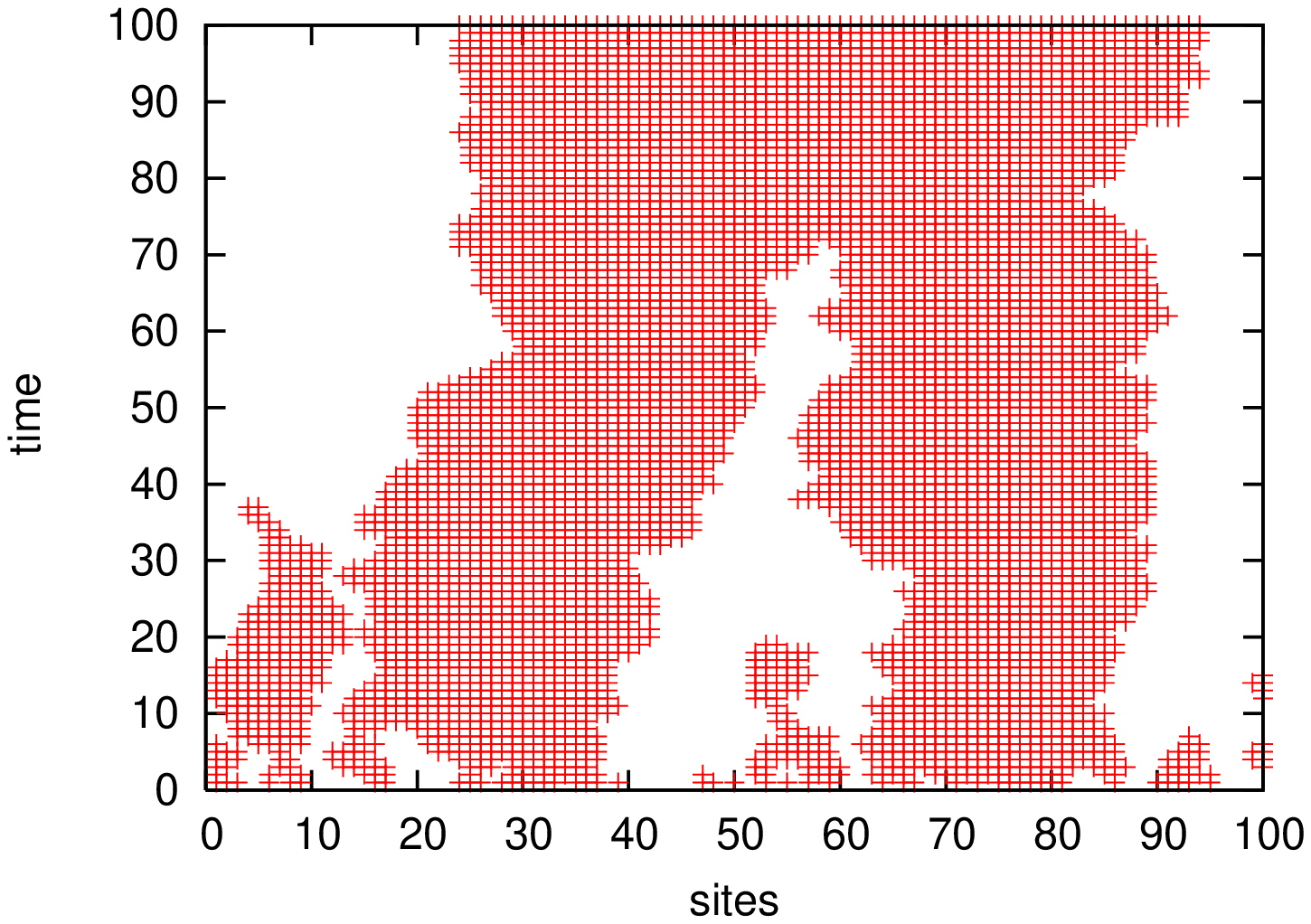}}}
\rotatebox{270}{\resizebox*{4.5cm}{!}{\includegraphics{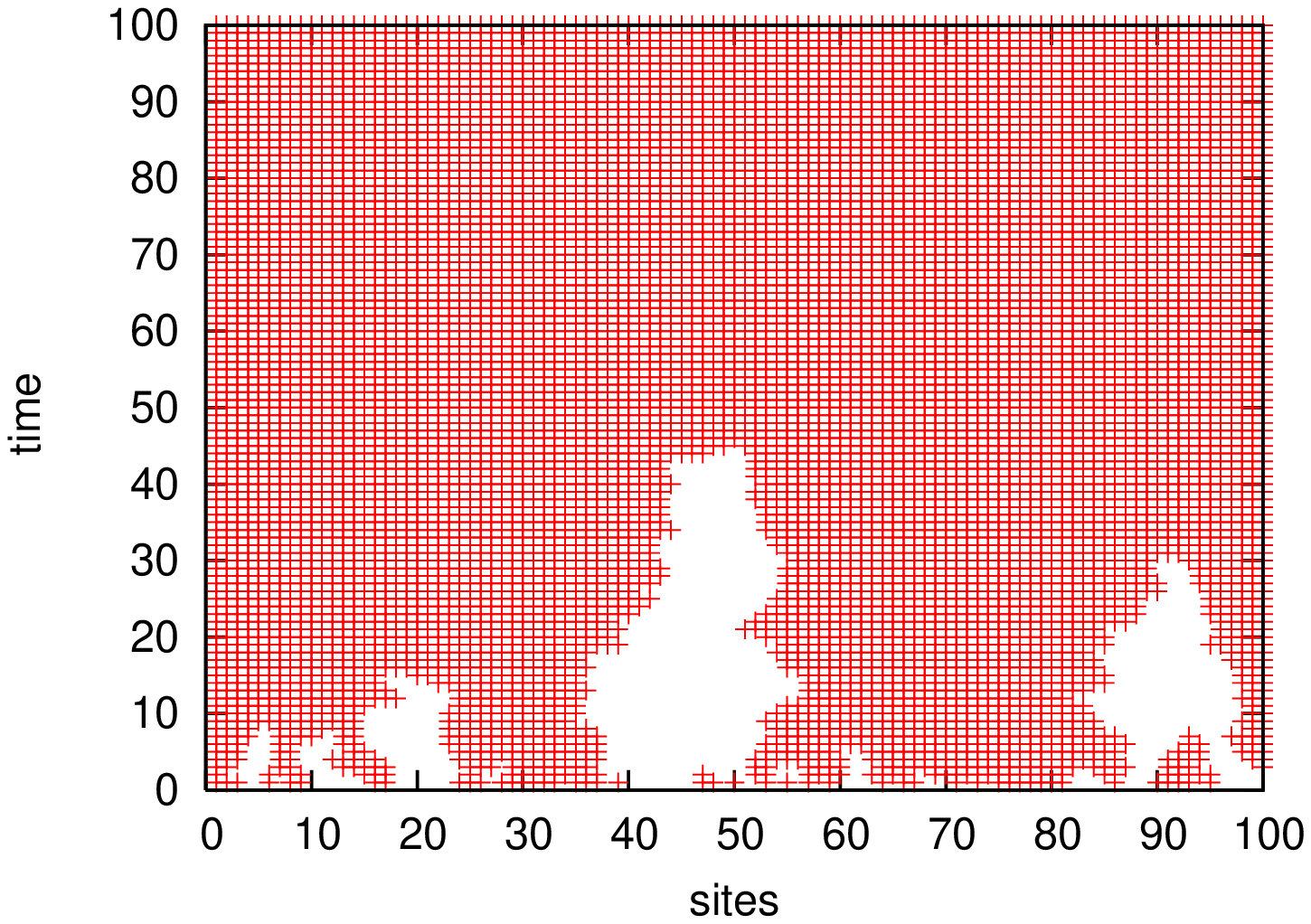}}}
\rotatebox{270}{\resizebox*{4.5cm}{!}{\includegraphics{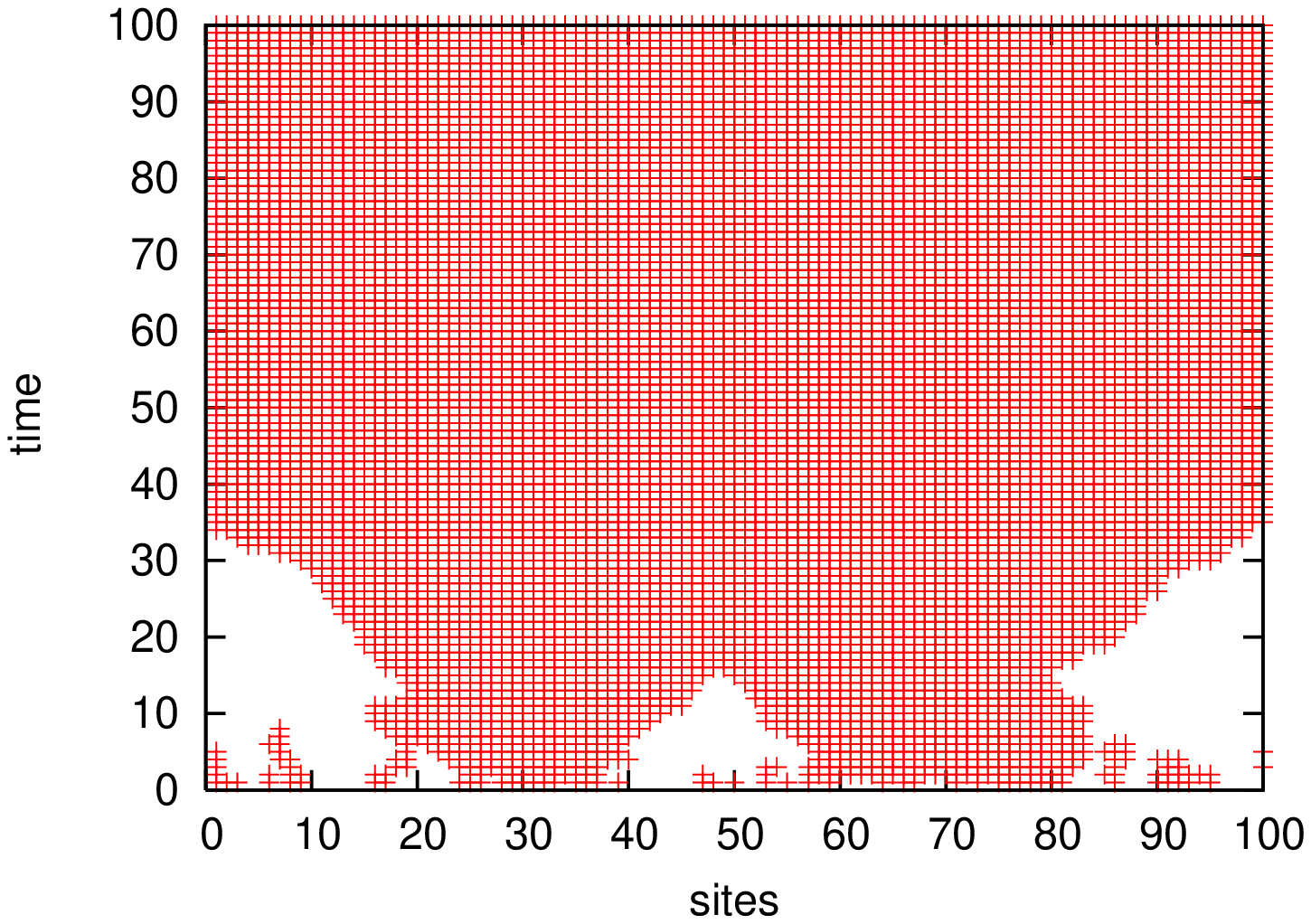}}}
\rotatebox{270}{\resizebox*{4.5cm}{!}{\includegraphics{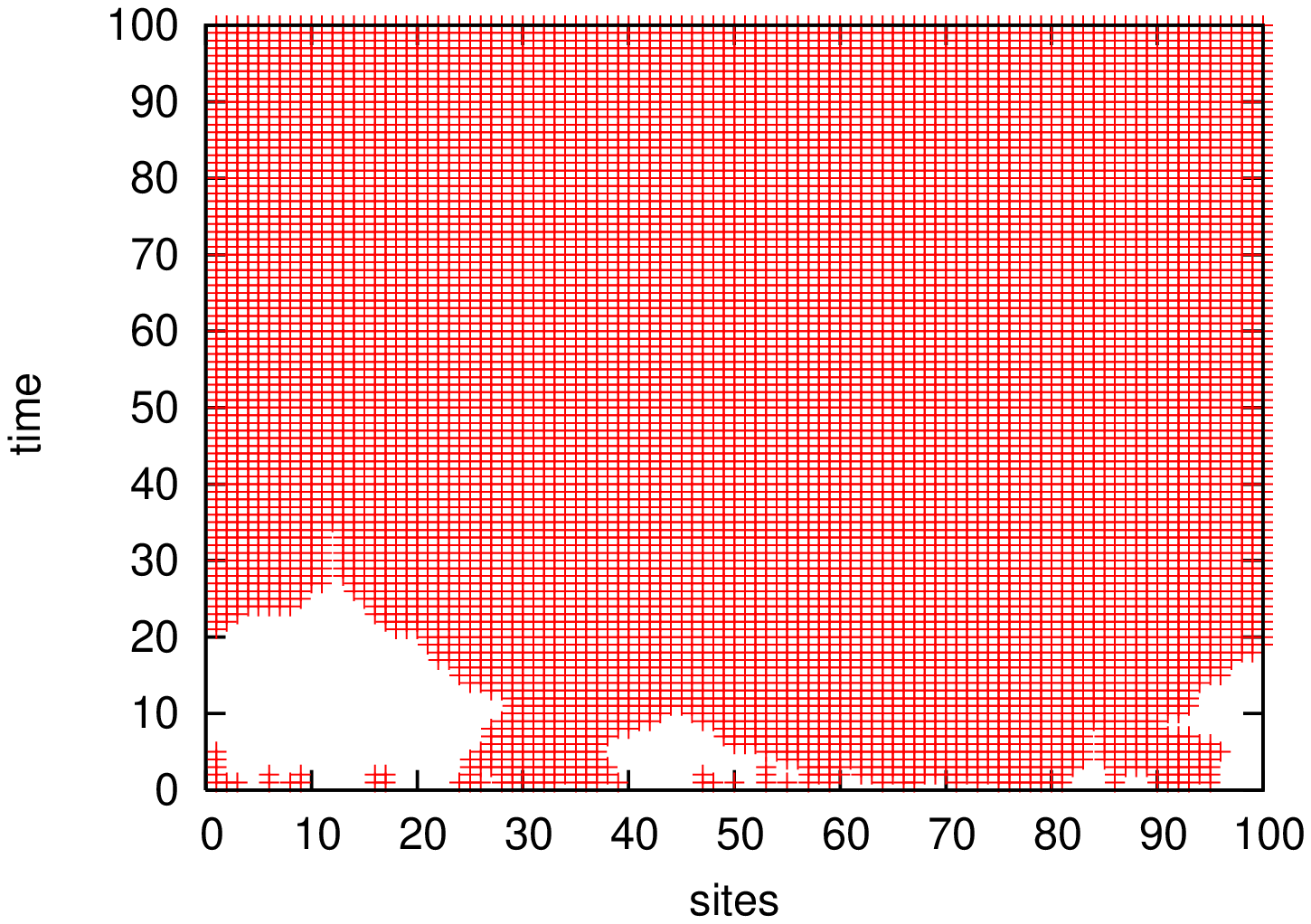}}}
\caption{(Color online) Snapshots of the system in time for different values of $\beta = 0.0, 0.005, 0.1$ and $\beta \to \infty$ (top to bottom) 
 showing that for any non-zero $\beta$, the system equilibriates
towards a homogeneous state much faster compared to the $\beta = 0 $ case. These snapshots are for a 
$L=100$ system.}
\end{figure}

In the simulations, we have generated systems of size $1000 \leq L \leq 10000$ with a mininum of 1000 initial 
configurations for the maximum size in general.  Only for $\beta =0$, the Ising limit, in which case 
 the   time taken to reach equilibrium is  order of magnitude higher than that 
for any nonzero $\beta$, smaller systems  have been simulated  in some calculations. 
Depending on the system size and time to equilibriate, maximum iteration times have been
set. Random sequential updating process has been used to control the spin flips.

On introducing $\beta$, we notice that well away from the Ising limit $\beta=0$, the dynamics gives $z \simeq 1.0$ and $ \theta \simeq  0.235 $ as in model I. However, as $\beta$ is made less than $O(10^{-1})$,
the behaviour of the relevant dynamic quantities deviate from a simple
power law behaviour. 
For example, the magnetisation shows an initial slow variation with time
followed by a rapid growth before reaching saturation for values of $\beta < 0.1$ (Fig. 2). It is difficult to
fit a  power law  in either regime.
This is true for  the domain wall fraction decay as well (not shown). 
In fact, the rapid growth of magnetisation at later times is apparently even faster than 
$t^{1/2}$, that obtained for model I (e.g., for $\beta = 0.001$). From the  snapshots of the system
for $\beta$ very close to zero,  it is seen that for the first few steps the system has a behaviour similar to the Ising model ($\beta =0$). This explains the 
slow growth of magnetisation initially. However, as soon as a domain  shrinks in size compared to 
its adjacent one, any non-zero $\beta$ makes it vanish very rapidly. 
However, it will be wrong to infer that the coersening process takes place faster 
than  in model I, because in comparison, in model I, 
the system equilibriates in times much lesser  than that for any finite $\beta$
(see Fig. 1).

\begin{figure}
\rotatebox{270}{\resizebox*{5cm}{!}{\includegraphics{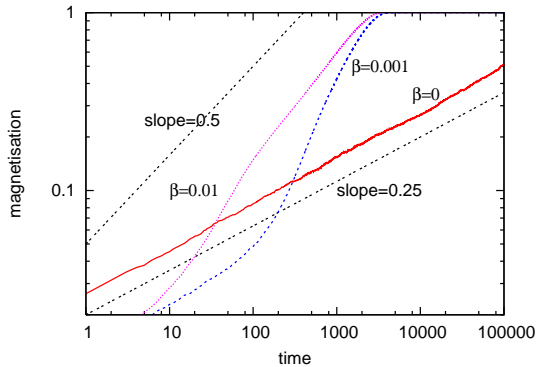}}}
\caption{(Color online) Magnetisation as a function of time is shown for 
$\beta = 0, 0.001$ and $0.01$. The two starightlines in the log
log plot have slopes corresponding to model I (0.5) and Ising model (0.25).
The $\beta =0$ result is for $L=2000$ while the others are for $L=10000$.}
\end{figure}

The question remains therefore whether and how  one can obtain an estimate of $z$
for $\beta \to 0$. Since a direct fitting fails, we try an indirect method. 
The average time $t_{eq}$ to reach the equilibrium state can be estimated from the 
time the magnetisation reaches a value unity. $t_{eq}$ is shown to scale  as $L^z$ in Ising model with $z=2$ and for 
$\beta \to \infty$,  $t_{eq}$ scales as $L^z$ with $z=1$ (inset of Fig. 3). Hence 
we plot $t_{eq}/L$ against $\beta$ for different $L$ and find an interesting
result. For  values of $\beta$ greater than 0.01, it shows a nice collapse, indicating 
$z =1$ here. As $\beta $ decreases, the deviation from a collapse starts appearing, it getting more pronounced for smaller values of $\beta$. However, at the same 
time, we notice that 
the deviation from a scaling 
$t_{eq} \propto L$  decreases for larger values of $L$ suggesting that 
the collapse as $\beta \to  0$ will improve with the system size.
Thus we conclude that the exponent $z$  equals  unity in the thermodynamic limit for 
any non-zero value of $\beta$. The deviations from the scaling  as 
$\beta \to 0$ 
is simply a finite size effect. 

\begin{figure}
\rotatebox{0}{\resizebox*{7cm}{!}{\includegraphics{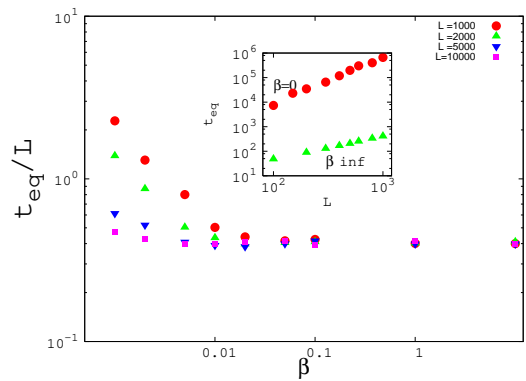}}}
\caption{(Color online) The values of the time to equilibriate $t_{eq}$ scaled by the sytem size shows that for large $\beta$ there is a nice collapse. For small $\beta$, there
are deviations from the collapse which decrease with the system size. Inset shows that $t_{eq}$ scales as $L^z$ for the limiting values $\beta =0$ with $z=2$ and for $\beta \to \infty$, $ z=1$.}
\end{figure}

Hence from the above behaviour we conclude that the model I behaviour 
is valid for any finite $\beta$ and a dynamic transition takes place exactly at $\beta = 0$.

Next we focus on the persistence data.
Once again, as $\beta \to 0$, it is difficult to fit a unique power law 
 to the persistence probability (Fig. 4). Here, in consistency with the 
magnetisation results, we find an initial decay of persistence quite fast and a late variation comparatively 
slower. The initial  variation can be fitted to a power law and  an estimate of $\theta$ made this way 
shows a tendency to continuously vary
towards the  $\beta=0$ value, i.e., 0.375. However, $\theta$ is not to be  obtained from the early  time behaviour 
 and
there is definitely a crossover to a different behaviour in later times before
the persistence reaches saturation.
Therefore determining $\theta$ from the initial variation is not a correct approach.

We even try  to obtain a collapse by plotting $P(t)/t^{-\theta}$ against ${L/t^z}$ using trial values of $z$ and $\theta$  as in \cite{bcs,biswas-sen}, 
but for $\beta \to 0$,
no collapse for large $L/t^z$, i.e., for small $t$, can be obtained, confirming once again that the determination of $\theta$  is 
not possible in  a straightforward manner.

We next try to find out whether the scaling law $P_{sat}(L) \propto L^{-\alpha}$ is valid for finite values of $\beta$. When we plot
 the saturation values of persistence against different system sizes, we do find nice power law fittings and hence estimates of $\alpha$ can be made (Fig. 5). 
We find   that $\alpha$ varies  between $0.22$  and $0.23$ with no systematics indicating
that it is independent of $\beta$. This once again supports the fact that
there is a transition at $\beta =0$ as $\alpha$ has a  known value (0.75) 
much larger for $\beta =0$.  
\begin{figure}
\rotatebox{270}{\resizebox*{6cm}{!}{\includegraphics{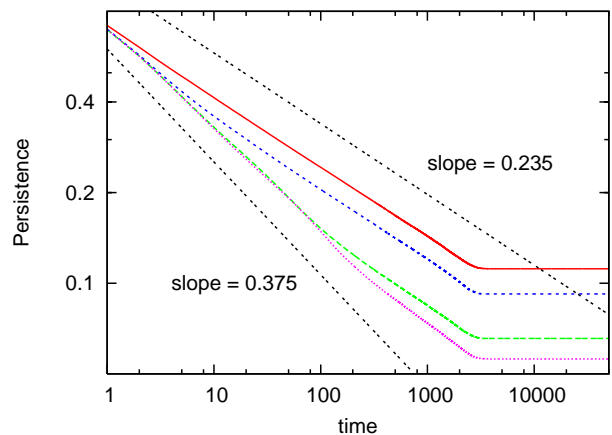}}}
\caption{(Color online) The persistence probability against time is shown for different values of $\beta$ ($\beta = 100,0.1, 0.01$ and $0.005$ (from top
to bottom); for
small $\beta$ the slope cannot be uniquely determined. The two straightlines 
in the log log plot have slopes corresponding to the model I (0.235) and Ising model (0.375).}
\end{figure}

Although $\alpha$ shows no dependence on $\beta$, the saturation values of the 
persistence probability show an
interesting dependence on $\beta$: for $\beta >>1$, it is independent of 
$\beta$ while for small values of 
$\beta$ it has a power law variation. We in fact find that the scaled variable $P_{sat}/{(\frac{\beta}{L}})^{\alpha}$
with $\alpha = 0.225$ shows a collapse when plotted against $\beta$ suggesting a scaling form:
\begin{equation}
P_{sat}(L, \beta) = (\beta/L)^{\alpha} f(\beta).
\end{equation}
The fact that for large values of $\beta$, the saturation values are independent
of $\beta$ suggests that $f(\beta)$ varies as $\beta ^{-\alpha}$ here.
We indeed find this kind of a behaviour with $f(\beta) = constant$ for $\beta << 1$ and  
$f(\beta) \propto \beta^{-\alpha}$ for $\beta >>1$ (see inset of Fig. 5).

We thus find that the effect of  the noise parameter $\beta$ 
is to cause a dynamic phase transition at $\beta=0$ showing that 
the behaviour of model I is indeed very robust. On the other hand, with respect to the Ising model, although the effect of noise is
not comparable to thermal fluctuations as far as order-disorder transitions are concerned, it does induce a dynamic phase
transition at $\beta =0$. The signature of the dynamical phase transition is seen in the variation of the 
dynamical quantities as the $\beta =0$  point is approached, there are also strong finite size effects. 

\begin{figure}
\rotatebox{0}{\resizebox*{8cm}{!}{\includegraphics{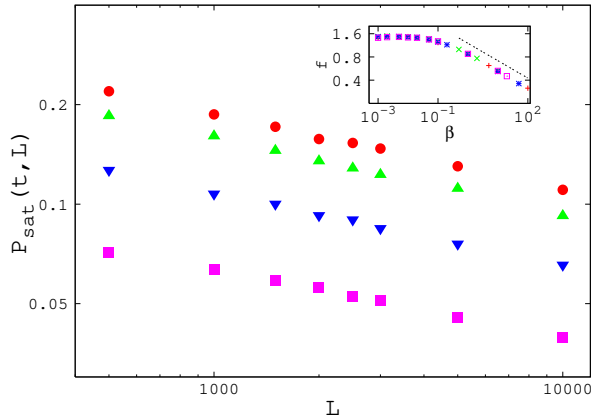}}}
\caption{(Color online) The saturation values of the persistene probability shows a
variation $L^{-\alpha}$ for  values of $\beta = 10.0, 0.1,0.01$ and $0.001$ (from top to bottom) with
$\alpha = 0.225$. Inset
shows that the scaled saturation values $P_{sat}/(\beta/L)^{\alpha} = f(\beta)$
 varies as $\beta ^{-\alpha}$
for large $\beta$.}
\end{figure}

One may raise the question as to what happens if $\beta $ is made negative. As expected, 
the system goes to a disordered state for any non-zero $\beta$ accompanied with 
exponential decay of persistence probability. 
In the spin picture, a negative value of  $\beta$ does not correspond to any physical model, but
in terms of domain wall movement, one has a system of mutually repulsive random walkers 
when $\beta < 0$. 
The random walkers tend to move away from their nearest neighbours
and therefore  cannot annihilate
each other but remain mobile all the time destroying the persistence
of the spins. Such situations was seen to arise in  spin systems like the ANNNI model \cite{annni} also.  
The dynamic behaviour is therefore different for $\beta < 0, \beta =0 $ and $\beta > 0$.
So, allowing negative values of $\beta$, one may say that there  is a dynamic phase
transition occurring at $\beta =0$ separating {\it {three}} different dynamical phases.

Acknowledgment: Financial supports from DST grant no. SR-S2/CMP-56/2007   and
discussions with P. Ray  and S. Biswas are acknowledged.


\begin{thebibliography}{99}
\bibitem{biswas-sen} S. Biswas and P. Sen,
Phys Rev E {\bf 80} 027101 (2009). 
\bibitem{absorb} 
H. Hinrichsen, Adv. Phys. {\bf {49}} 815 (2000); J. Marro and R. Dickman,
{\it { Nonequilibrium Phase Transitions in Lattice Models}} (Cambridge 
University Press, Cambridge), 1999; G. Odor, Rev. Mod. Phys. {\bf{76}} 663 (2004).
\bibitem{vote} T. M. Liggett   \textit{Interacting Particle Systems: Contact, Voter and Exclusion Processes} (Springer-Verlag Berlin 1999).
\bibitem{opinion1} D. Stauffer, in {\it {Encyclopedia of Complexity 
and Systems Science}} edited by R. A. Meyers (Springer, New York, 2009);
 K. Sznajd-Weron and  J. Sznajd,   Int. J. Mod. Phys C {\bf{11}} 1157 (2000);
S. Galam, Int. J. Mod. Phys. C {\bf {19}} 409 (2008).
\bibitem{opinion2} A. Baronchelli, L. Dall'Asta, A. Barrat, and V. Loreto,
Phys. Rev. E {\bf{76}}, 051102 (2007);
C. Castellano, M. Marsili and A. Vespignani,
Phys. Rev. Lett. {\bf{85}} 3536 (2000).
\bibitem{hinrich2} J. Fuchs, J. Schelter, F. Ginelli and  H. Hinrichsen,
J. Stat. Mech. P04015 (2008).
\bibitem{stauffer2} D. Stauffer and P. M. C. de Oliveira, Eur. Phys. J B
{\bf{30}} 587 (2002.)
\bibitem{sanchez} J. R. Sanchez, arXiv:cond-mat/{{0408518v1}}
\bibitem{shukla} P. Shukla, J. Phys. A Math. Gen. {\bf{38}} 5441 (2005).
\bibitem{derrida} 
B. Derrida, A. J. Bray and C. Godreche, {J.Phys. A {\bf{27}}
L357 (1994)}. 
\bibitem{bray} A. J. Bray, {Adv. Phys. {\bf{43}} 357 (1994) and the references
therein}.
For a review on persistence, see S. N. Majumdar, {Curr. Sci.
{\bf{77}} 370 (1999)}.
\bibitem{pray} G. Manoj and P. Ray, Phys. Rev. E {\bf {62}} 7755 (2000); G. Manoj and P. Ray, J. Phys A {\bf{33}} 5489 (2000).
\bibitem{bcs} S. Biswas, A. K. Chandra and P. Sen, Phys. Rev. E {\bf{78}}, 041119 (2008)
\bibitem{annni}
P. K. Das, S. Dasgupta and P. Sen, J. Phys. A: Math. Theor. {\bf{40}}  6013 (2007).

 \end{thebibliography}
\end{document}